\documentclass[sigconf,authorversion,nonacm]{acmart}
\AtBeginDocument{%
  }

\usepackage{graphics}  
\usepackage{subcaption}
\usepackage{enumitem}
\usepackage{csquotes}
\usepackage{multirow}
\usepackage{tabularx}
\usepackage{tikz}
\renewcommand{\mkbegdispquote}[2]{\itshape}
\usepackage{cleveref}[2012/02/15]
\crefformat{footnote}{#2\footnotemark[#1]#3}
\usepackage{xcolor}

\begin{document}

\title[Student-AI Collaborative Hint Writing]{Hint-Writing with Deferred AI Assistance: Fostering Critical Engagement in Data Science Education}

\author{Anjali Singh}
\email{anjali.singh@ischool.utexas.edu}
\affiliation{%
  \institution{The University of Texas at Austin}
  \country{USA}
}

\author{Christopher Brooks}
\email{brooksch@umich.edu}
\affiliation{%
  \institution{University of Michigan}
  \country{USA}
}

\author{Warren Li}
\email{warren@ld-insights.com}
\affiliation{%
  \institution{Learning Data Insights}
  \country{USA}
}

\author{Juho Kim}
\email{juhokim@kaist.ac.kr}
\affiliation{%
  \institution{KAIST}
  \country{Republic of Korea}
}

\author{Xu Wang}
\email{xwanghci@umich.edu}
\affiliation{%
  \institution{University of Michigan}
  \country{USA}
}

\renewcommand{\shortauthors}{Singh et al.}

\begin{abstract}
Generating hints for incorrect code is a cognitively demanding task that fosters learning and metacognitive development. 
This study investigates three designs for personalized, scalable, and reflective hint-writing activities within a data science course: (i) writing a hint independently, (ii) writing a hint with on-demand AI assistance, and (iii) \textit{deferred AI assistance}, in which students first write a hint independently and then revise it with the help of an AI-generated one. We examine how AI support can scaffold the learning process without diminishing students’ productive cognitive effort.
Through a randomized controlled experiment with graduate-level students ($N=97$), we found that deferring AI assistance
leads to the highest-quality hints. Further, this design helps students identify a wide range of mistakes they otherwise struggle to identify without any AI assistance. Students valued these activities as opportunities to practice debugging and critically engage with AI outputs---skills that are now critical for learners to acquire as programming becomes increasingly automated and the use of AI for learning grows. Our findings also highlight key considerations for designing student-AI collaborative learning experiences to sustain student engagement, maintain appropriate cognitive load, and mitigate negative effects of AI, such as introducing redundancies and extraneous information into student work. 
\end{abstract}



\keywords{Hint Writing, Deferring AI Assistance, Data Science Education}

\settopmatter{printacmref=false}
\setcopyright{none}
\renewcommand\footnotetextcopyrightpermission[1]{}
\pagestyle{plain}


\maketitle

\section{Introduction}
As AI becomes increasingly capable of undertaking complex tasks once performed by humans, such as programming and writing, higher education must prepare students to develop essential skills such as AI literacy, metacognition, and critical and reflective thinking \cite{atchley2024human}. There are also growing concerns that students may use AI as a crutch rather than a scaffold, bypassing the effortful cognitive processes that drive durable learning \cite{fan2024beware, gerlich2025ai,zhai2024effects}. These challenges call for deliberate pedagogical interventions that foster reflective thinking \cite{dewey2022we} and meaningful student-AI interaction, helping learners engage critically with AI and develop the evaluative judgment skills that are increasingly important in the age of AI  \cite{bearman2024developing, uanachain2025generative}.
This need is particularly acute in domains such as computing and data science, where students can use AI to generate working code without fully developing foundational competencies such as computational thinking, data literacy, and semantic understanding of programming libraries \cite{denny2024computing, singh2024investigating}.

We developed personalized, reflective assignments that required writing hints for incorrect programming assignment solutions. These assignments were developed to involve students in practicing debugging of incorrect data science code while also engaging them in critical thinking, communication, and reflection through hint writing \citep{harris2013opportunities, nicol2014rethinking}. To make hint-writing personalized and scalable in large courses, we designed tasks where students compared their own correct (or instructor-provided) solutions with algorithmically selected similar but incorrect solutions, following which they were prompted to write a hint for the incorrect solution (see Figure \ref{fig:hint_writing_designs}).
As hint-writing is an inherently effortful process, requiring both cognitive and metacognitive engagement \cite{endres2024constructive, fukaya2013explanation}, it provides rich context for exploring how students can engage with AI in constructive ways for complex problem-solving. Large Language Models (LLMs) offer the potential to support hint-writing by helping students frame hints and identify errors. For instance, recent work suggests that LLM-generated hints can provide valuable scaffolding in hint-writing activities by asking students to revise and improve upon LLM-generated hints; however, this approach can lead to over-reliance, with learners adopting AI suggestions uncritically rather than exercising independent judgment \cite{singh2023bridging}.

A promising strategy to mitigate overreliance is \textit{deferring} AI assistance---providing the assistance only after students have first attempted the task independently. Research shows that deferring assistance can strengthen metacognition by allowing students to confront the difficulty of a problem before receiving support \cite{fisher2021harder}. This approach also aligns with the ``Assistance Dilemma'' \cite{koedinger2015data}, which highlights the importance of balancing explicit guidance with opportunities for productive struggle \cite{murdoch2020feeling} to foster deeper learning. According to this dilemma, while AI assistance can enhance learning, excessive assistance risks undermining learners' capacity for critical thinking and independent learning, whereas insufficient assistance may make tasks too challenging. Beyond deferral of AI assistance, allowing students to revise and improve their independently generated work after receiving AI support can create further opportunities for reflection and critique. 

To examine the impact of deferring AI assistance relative to no assistance (reflecting traditional approaches) and on-demand AI assistance (reflecting typical availability of AI support), we compared three approaches to hint writing: \textit{Baseline}, \textit{AI-assistance}, and \textit{Deferred AI-assistance} (Figure \ref{fig:hint_writing_designs}). In the \textit{Baseline} condition, students received no AI assistance and wrote a hint independently. In the \textit{AI-assistance} condition, students had on-demand access to an LLM-generated hint for the same incorrect solution for which they needed to write a hint. In the \textit{Deferred AI-assistance} condition, students first attempted to write their own hint and then received an LLM-generated hint, following which they were prompted to rewrite their hint. Comparing these three designs allows us to identify whether, and at what stage, AI assistance should be integrated into a valuable yet complex learning activity like hint writing, as well as the considerations involved in developing pedagogical approaches that foster students’ critical engagement with AI. Notably, while all three designs support learners in practicing code evaluation, the designs with AI-assistance additionally foster evaluative judgment of AI-generated output.

This work reports findings from a study conducted in an online graduate-level data science course (N=97) at a large public university in the U.S., in which students completed two hint-writing assignments under one of three design conditions. This study investigates how the different designs influenced students’ learning outcomes, time on task, hint quality, perceptions of the task’s learning benefits, and their interactions with AI-generated hints. Specifically, we address the following research questions:

\begin{figure*}[t]
\centering
\subfloat{%
  \includegraphics[clip,width=\textwidth]{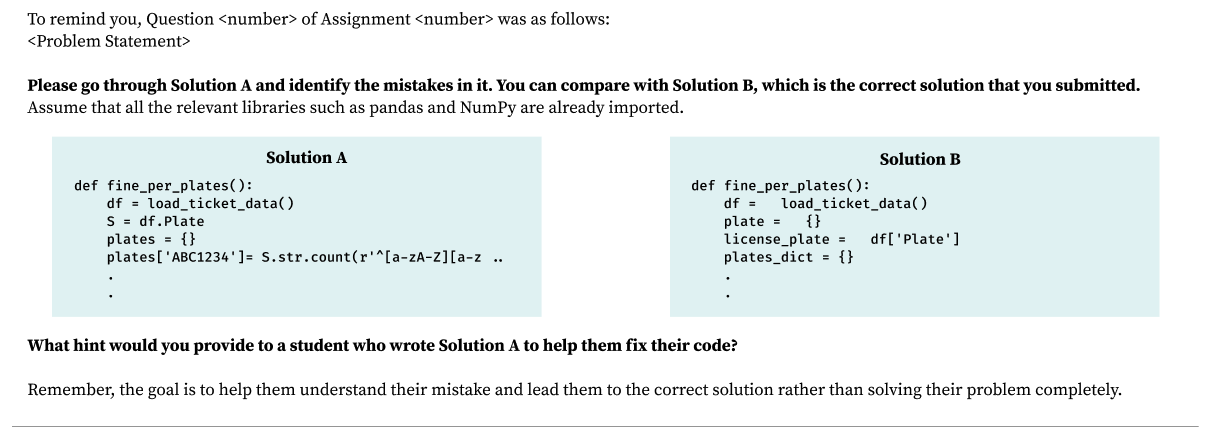}%
}
\\
  \vspace{-4mm}
\subfloat{%
  \includegraphics[clip,width=\textwidth]{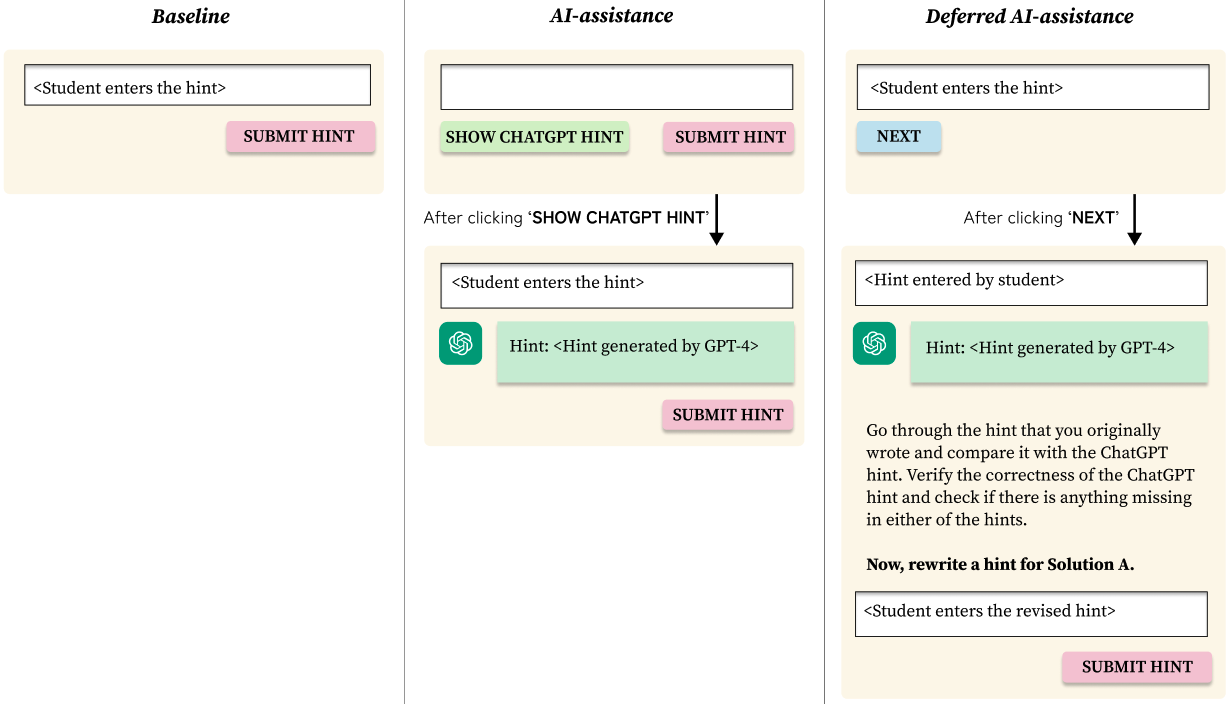}%
}
  \caption{Interface designs for the three hint writing conditions:
  First, students in all conditions were reminded of the question from the latest programming assignment for which they would write a hint. Next, they were shown a correct (Solution B) and an incorrect (Solution A) solution side by side. Solution B was a student's own solution if it was correct, else it was the instructor's solution. Solution A was selected algorithmically to be similarity to Solution B. Finally, students were asked to write a hint for Solution A as follows: 
  (i) students in the \textit{Baseline} condition wrote hints on their own, (ii) students in the \textit{AI-assistance} condition wrote a hint with the option to view an AI-generated hint by clicking on the `SHOW CHATGPT HINT' button, and (iii) students in the \textit{Deferred AI-assistance} condition first wrote a hint independently, then read an AI-generated hint and finally revised their hint.}
  ~\label{fig:hint_writing_designs}
  \vspace{-4mm}
\end{figure*}

\begin{itemize}
\item \textbf{RQ1}: How does each hint-writing design impact students' learning outcomes?
 \item \textbf{RQ2}:  How does each hint-writing design impact the quality of student-generated hints?
 \item \textbf{RQ3}: What are students' perceptions about the pedagogical value of each hint writing design? 
\end{itemize}

Our findings show that deferring AI assistance led to the highest quality hints. Students in the \textit{Deferred AI-assistance} condition were significantly more successful at identifying mistakes in the incorrect solutions provided to them compared to those in the \textit{Baseline} condition, while no significant differences were observed between the other two pairs of conditions. However, despite these benefits of deferring AI assistance, it did not fully offset the negative effects due to redundancies and inaccuracies in AI-generated hints, underscoring the substantial influence AI outputs can exert on students’ reasoning processes. Students in the \textit{Deferred AI-assistance} condition also spent significantly more time on the first hint-writing activity---more than twice as long on average compared to the other two conditions, demonstrating greater engagement--- although this difference disappeared in the second week. Across conditions, we found no significant differences in the learning outcomes.

Based on these findings, we offer practical guidance for designing hint-writing learning experiences that foster meaningful, critical engagement with AI, contributing to the growing body of work on strategies for integrating AI in ways that preserve cognitive effort and promote active learning \cite{dai2023reconceptualizing, kazemitabaar2025exploring, kumar2024guiding, oates2025chatgpt, park2024promise, prather2024widening}.
While much research has explored the use of LLMs for personalized tutoring, this work contributes to the limited research on using LLM-based support---despite its limitations---not only as feedback but also as an object of student critique.
Beyond empirical insights, we also contribute a dataset\cref{main_footnote} of student and AI-generated hints, paired with correct and incorrect programs, along with their evaluations based on a validated rubric. As research on feedback generation with LLMs expands \cite{azaiz2024feedback, estevez2024evaluation, gabbay2024combining, kiesler2023exploring, koutcheme2022towards, phung2024automating, stamper2024enhancing}, our dataset and rubric provide theoretically grounded resources for advancing automated evaluation and the generation of high-quality feedback in data science courses.

\section{Related Work}

\subsubsection*{\textbf{Learning from Mistakes and Writing Hints}}
The first part of the hint writing activity involved contrasting a correct data science program with an erroneous one. Prior research has shown that contrasting incorrect methods with correct ones enhances learning by directing students' attention to distinguishing features of the correct method and the underlying concepts \cite{durkin2012effectiveness}. It also promotes procedural flexibility, i.e., the ability to understand, apply, and compare multiple problem-solving strategies \cite{rittle2011power}. Reviewing incorrect solutions also offers students the opportunity to see examples of potential errors and understand their consequences \cite{cho2011learning, kaufman2011students}.

The next part of the task involved writing a hint for the erroneous program, which involves three key steps: (i) identifying the mistakes within the incorrect solution, (ii) linking those mistakes to underlying misconceptions, and (iii) offering suggestions for correcting the mistakes without providing the complete solution, as doing so would undermine the hint recipient's knowledge construction \cite{singh2023bridging}. The first two steps engage students in retrieval practice, which improves long-term retention by recalling relevant facts and concepts from memory \cite{roediger2011critical}. Together with elaboration and explanation, which are required in the third step, these steps contribute to improving metacognitive abilities \cite{endres2024constructive}. The third step involves \textit{explaining} the mistakes and how to fix them. This enhances students' comprehension by promoting deep-level cognitive processes that involve organization and integration of information \cite{fiorella2016eight}. It also helps elicit metacognitive processes that promote the deployment of effective cognitive strategies \cite{fukaya2013explanation}. 
Additionally, in data science,
actively involving students in writing hints for incorrect solutions by comparing different solutions to the same problem can be advantageous for exposing them to the diverse problem-solving approaches \cite{nguyen2021exploring}.  Lastly, as programming becomes increasingly automated with the advent of AI code generators, engaging students in hint writing helps them practice skills such as debugging and code comprehension, which are critical for evaluating and effectively using AI-generated code \cite{kazemitabaar2023studying}. 


\subsubsection*{\textbf{Rethinking Pedagogical Approaches in the Age of AI}}
The rapid integration of AI into learning environments necessitates rethinking the skills students must acquire, and accordingly the pedagogical approaches that need to be developed to support them in acquiring those skills  \cite{dai2023reconceptualizing, long2020ai,park2024promise, uanachain2025generative, yan2024promises}.
In addition to domain-specific expertise, competencies such as critical thinking and AI literacy have become essential \cite{atchley2024human, long2020ai,park2024promise}. There are growing concerns regarding students' over-reliance on AI, particularly in programming education, where excessive dependence on LLMs for code generation can impede the development of computational thinking and metacognitive skills \cite{denny2024computing} and lead to a decrease in student grades \cite{jovst2024impact}. Studies have shown that support from AI code-generators \cite{kazemitabaar2023novices} or AI-generated feedback \cite{pankiewicz2023large} can contribute to such over-reliance, and students struggle to perform as well in the absence of AI assistance \cite{darvishi2024impact}. 
These findings are linked to cognitive offloading---when students delegate core cognitive tasks to AI at the expense of self-regulation and deeper engagement \cite{risko2016cognitive, fan2024beware}.

In response, researchers have begun exploring curricular and assessment designs that encourage students to engage critically with AI outputs while strengthening essential 21st-century skills such as critical thinking and communication. Some approaches introduce structured friction into student-AI interactions to promote active engagement through novel user interface designs \cite{kazemitabaar2025exploring}, while others involve students in critiquing and revising AI-generated outputs, such as essays \cite{oates2025chatgpt} or hints \cite{singh2023bridging}. An advantage of the latter approach is that it may foster lasting habits of critical engagement with AI beyond the classroom, unlike designs that deliberately add friction through unique interfaces, as such interactions may not transfer to commercial AI tools offering seamless, frictionless experiences \cite{chen2024exploring}. Yet, revision tasks still risk over-reliance as students have been found to adopt AI suggestions without fully exercising independent judgment \cite{singh2023bridging}. 

Deferring AI assistance, by requiring students to attempt a solution independently before accessing one that is generated by AI, can be helpful for inducing more independent thinking. Prior work shows that deferred assistance can strengthen metacognition by exposing learners to the difficulty of the task \cite{fisher2021harder}. Further, as overreliance on AI increases with trust in AI outputs \cite{gerlich2025ai}, prompting students to independently solve a problem before seeing the AI solution can help calibrate their trust in AI, which can lower their reliance. This study builds on these insights to investigate deferred AI assistance as a design for student-AI collaborative hint writing, aiming to promote deeper engagement and more independent thinking in data science education.

\section{Method}
\subsection{Study Design}
\label{study_design}
\subsubsection*{\textbf{Educational Context}}
We conducted a study comparing three personalized hint writing assignment designs in the September 2023 offering of an online graduate-level Data Manipulation course at the University of Michigan. 
The course was taught over four weeks and had weekly programming assignments in Python that were automatically graded. Students were allowed unlimited submissions until the deadline for the programming assignments. They completed a pre-test at the beginning of the course, engaged in the hint writing assignments in two different weeks (Week 1 and Week 2), and took a post-test at the end of the course. We developed new interfaces that students were directed to for the hint writing assignments. The hint-writing activities---which involved reflecting on solutions to the most recent programming assignment, as described in the following subsection---were assigned immediately after the deadlines for the second and third programming assignments, similar to prior work \cite{singh2023bridging, glassman2016learnersourcing}. This ensured that students’ problem-solving approaches for those assignments remained fresh in their minds.

\begin{figure}[!b]
    \vspace{-5mm}
  \centering  \includegraphics[width=\columnwidth]{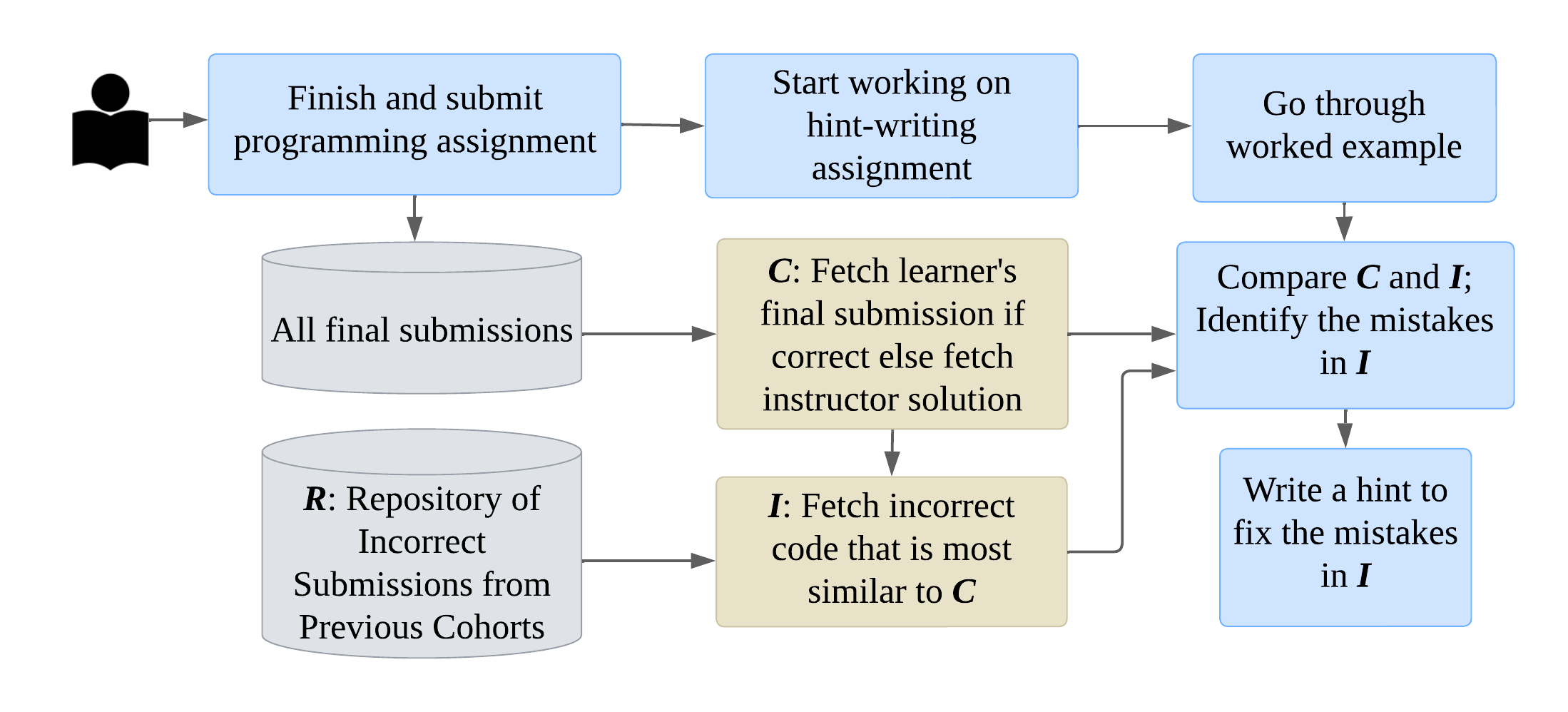}
  \caption{Workflow for creation of the personalized hint writing assignments.} 
    \label{full_pipeline}
\end{figure}

\begin{figure*}[t]
  \centering  \fbox{\includegraphics[width=\textwidth]{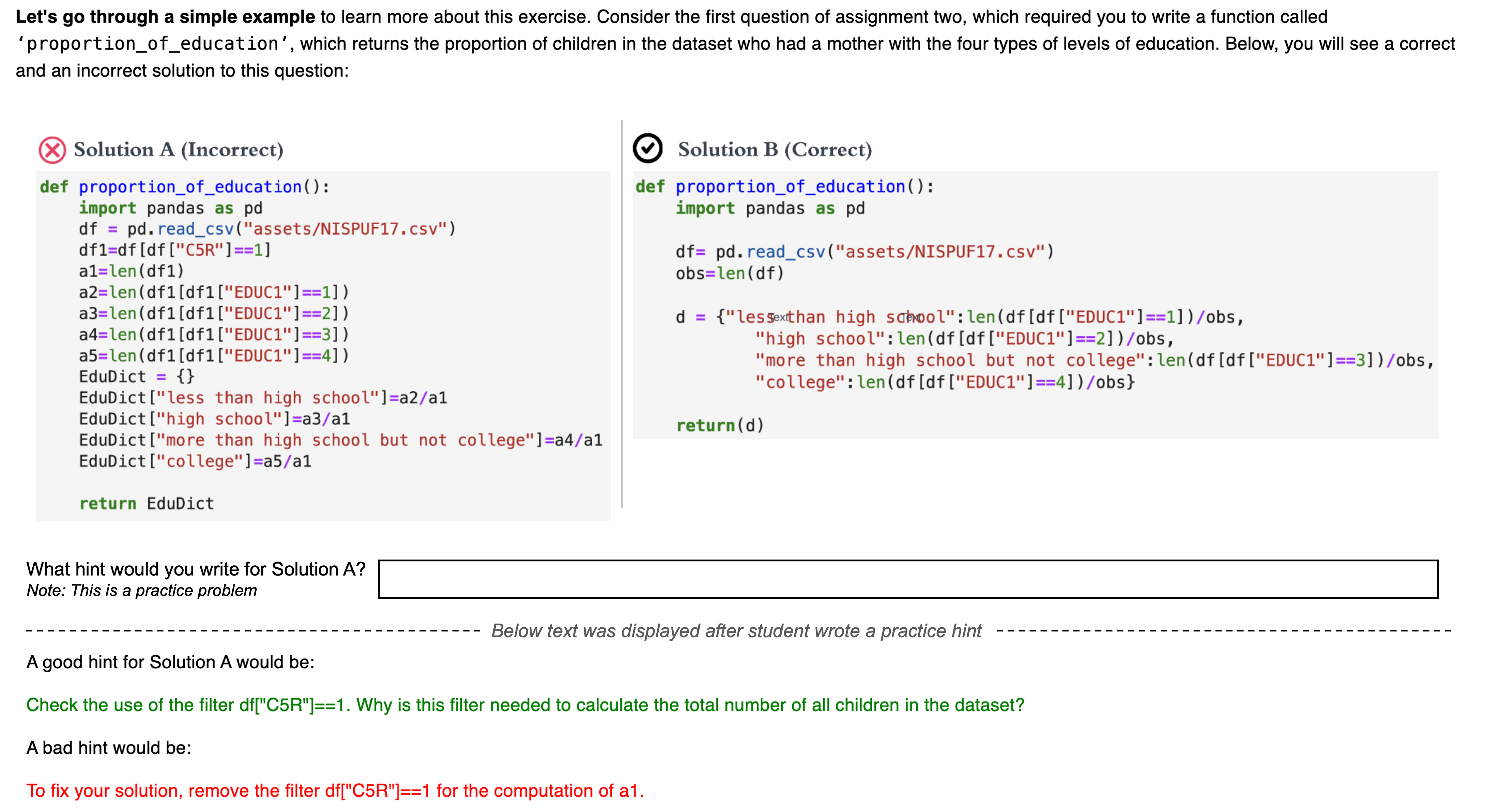}}
  \caption{Warm-up exercise demonstrating how to write a good hint.} 
    \label{fig:worked_example_505}
    \vspace{-2mm}
\end{figure*}

\subsubsection*{\textbf{Hint Writing Assignment Design}}
The landing page of the hint writing assignment\cref{main_footnote} explained the task goal---to write a hint that helps the hint receiver correct mistakes without revealing the full solution---and described the condition-specific procedure, e.g., those in the \textit{Deferred AI-assistance} condition were informed that they would first write a hint independently, then view a GPT-generated hint and then revise their original hint. The page also described the potential learning benefits of the exercise.
Next, they were provided tips to write a helpful hint. As a warm-up exercise, students were shown examples of a correct and an incorrect solution side by side and asked to write a hint for the incorrect solution, similar to the main task (Figure \ref{fig:worked_example_505}). After they wrote a hint, they were provided with examples of a good and bad hint for the same solution, to reinforce their understanding of the characteristics of good hints.
After the warm-up activity, students proceeded to the main task by going to the next page. The instructions for one of the problems (\texttt{$P$}) from the most recent programming assignment were reproduced (Figure \ref{fig:hint_writing_designs}). Students were then prompted to compare two solutions to that problem--a correct solution \texttt{$C$} and an incorrect solution \texttt{$I$}--displayed side by side. The method for determining \texttt{$C$} and \texttt{$I$} for each student is detailed in the next subsection. After comparing \texttt{$C$} and \texttt{$I$}, students were instructed to write a hint for \texttt{$I$} and were informed of the specific design of their hint-writing task, depending on the condition to which they were assigned.
As a small incentive, students were awarded points worth 5\% of the total course grade, solely on task completion, for each hint-writing assignment. To ensure that students do not use any outside AI assistance, they signed an honor code at the beginning of the course affirming that all submitted work would be their own. 

\subsubsection*{\textbf{Experimental Conditions}}
Students were randomly assigned to one of the three conditions
(Figure \ref{fig:hint_writing_designs}). 
Students in the \textit{Baseline} condition wrote hints independently. Those in the \textit{AI-assistance} condition had the option to view an AI-generated hint for \texttt{$I$} at any point during the task by clicking on the ``Show ChatGPT Hint" button. 
Those in the \textit{Deferred AI-assistance} condition first wrote a hint independently. Then, an AI-generated hint for \texttt{$I$} was displayed, and students were asked to compare it with their independently written hint. Finally, they were asked to revise their original hint. 
Students in both the \textit{AI-assistance} and \textit{Deferred AI-assistance} conditions were informed that the AI-generated hints provided to them could be incorrect, incomplete, or both, and they would be responsible for validating their accuracy.

\subsubsection*{\textbf{Selection of Correct (\texttt{$C$}) and Incorrect (\texttt{$I$}) Solutions}} The selection of \texttt{$C$} and \texttt{$I$} was guided by two primary factors: (i) students should have the opportunity to reflect on their own work, and (ii) selecting similar artifacts for comparison can make the task more tractable \cite{choi2022algosolve,glassman2016learnersourcing, margulieux2019finding}, particularly in data science where multiple diverse approaches to solve a problem typically exist \cite{nguyen2021exploring}. 
Given these considerations, if a student's final submission for \texttt{$P$} was correct, it was used as the correct solution \texttt{$C$}; otherwise, the correct solution supplied by the instructor served as \texttt{$C$}. The incorrect solution \texttt{$I$} was selected from a repository \texttt{$R$} of incorrect solutions submitted by learners from previous course offerings. \texttt{$R$} consisted of 12 and 13 unique incorrect solutions of the second and third programming assignments, respectively, covering a wide range of approaches.

Given a given correct solution \texttt{$C'$}, the most similar incorrect solution \texttt{$I'$} was selected from \texttt{$R$} using a keyword-based similarity metric. We consider all the imported libraries and built-in functions in a program as keywords, which were extracted using the programs' Abstract Syntax Trees (ASTs) via the \texttt{ast} module in Python\footnote{\url{https://docs.python.org/3/library/ast.html}}.
The keyword-based similarity metric computes the Jaccard coefficient \citep{niwattanakul2013using} between the keywords of two programs. Figure \ref{full_pipeline} shows the workflow for creating the personalized hint-writing assignments.

\subsubsection*{\textbf{Generating Hints for AI-assistance}}
We prompted GPT-4 to generate the hints\cref{main_footnote} used for providing AI assistance. To observe how students make use of them or get biased by them, we intentionally allowed for hints of varying quality, and therefore used zero-shot prompting \cite{schulhoff2024prompt}.
Each GPT-generated hint (hereafter referred to as GPT-hint), along with the corresponding incorrect answer for which the hint was generated, was stored in a database to be retrieved as needed when students engaged in hint writing. 

\subsection{Data Collection and Analysis}
\label{sec:data_analysis}
\subsubsection*{\textbf{Learning Outcomes (RQ1)}}
As mentioned in Section \ref{study_design}, students were assigned pre- and post-tests. The pre-test consisted of 10 Multiple Choice Questions (MCQs), with each MCQ having a single correct answer and assessing introductory Python knowledge. 
The post-test consisted of 2 MCQs with a single correct answer (worth 1 point each) and 4 MCQs with more than one correct answer (worth 2 points each).
This test assessed students' knowledge of data manipulation, particularly debugging skills, which the students were expected to have practiced by engaging in the hint writing tasks. 
Due to the non-isomorphic nature of the pre- and post-tests, differences in learning outcomes between conditions were evaluated by comparing post-test scores while controlling for prior knowledge. 


\subsubsection*{\textbf{Quality of Hints (RQ2)}}

\begin{figure}
  \centering  
  \includegraphics[width=\columnwidth]{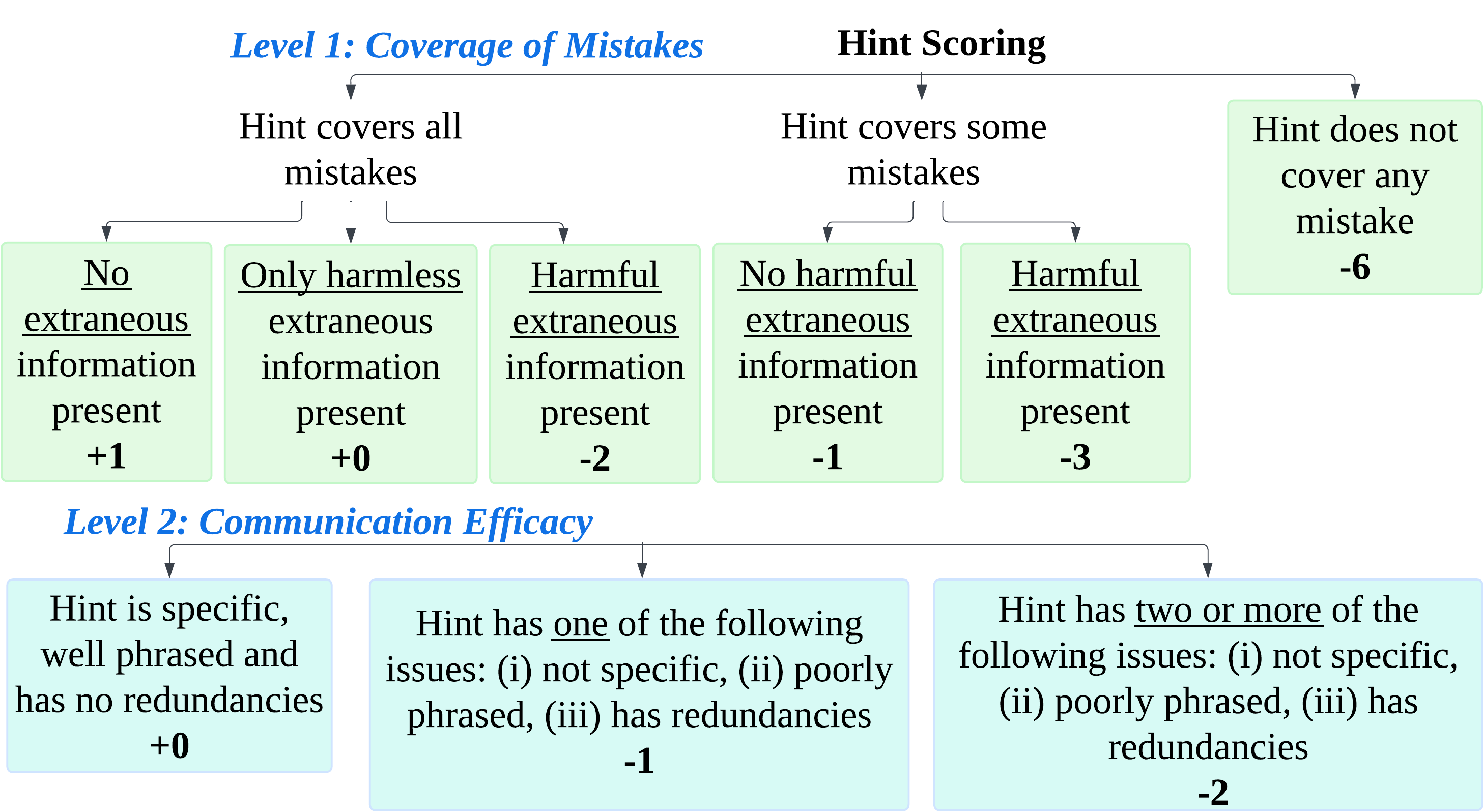}
  \caption{Hint scoring rubric for student-generated hints} 
  \label{fig:utility_levels}
  \vspace{-4mm}
\end{figure}

Building upon criteria used for evaluating hints in prior research \cite{marwan2021investigating, singh2023bridging, thurlings2013understanding}, two authors who had previously taught the course created a rubric for evaluating student-generated (the revised hint in the case of the \textit{Deferred-AI assistance} group) and AI-generated hints. The authors independently coded a random sample of 20\% of all hints using the rubric, discussed cases of disagreement, and engaged in three iterations on the definition of each criterion. The rubric\cref{main_footnote} was finalized after coding a new sample of 15\% of hints, for which a moderately high agreement of 0.78 Cohen's Kappa was achieved. Then, the first author used the rubric (Figure \ref{fig:utility_levels}) to rate all the hints.
The rubric consisted of the following dimensions adapted from prior research on effective forms of formative feedback \cite{marwan2021investigating, thurlings2013understanding}: \textit{accuracy}, \textit{specificity} (provides specific details about mistakes and suggestions to fix them without giving away the full solution), and \textit{phrasing} (use of positive and constructive language). We also added the \textit{lack of extraneous information} dimension based on recent work on AI-assisted hint writing \cite{singh2023bridging}. Any information in the hint that is not necessary for addressing the identified mistakes was considered as extraneous. Based on the iterative discussions to finalize the rubric, we categorized extraneous information as: \textit{harmful}, if the information is detrimental to students' learning and \textit{harmless}, if it is not detrimental to learning but not necessary to fix the mistakes, e.g., code formatting suggestions. Additionally, we added the dimension \textit{lack of redundancies} based on the observation that several GPT-hints consisted of redundant information, which can further impact the quality of student-generated hints.

As shown in Figure \ref{fig:utility_levels}, hint scoring was done at two levels: \textit{Level 1} evaluates the hint's \textit{coverage of mistakes}, i.e., to what extent the hint correctly identifies mistakes and excludes extraneous information; \textit{Level 2} evaluates the \textit{communication efficacy}, i.e., how effectively the hint provides guidance on how to address the identified mistakes based on specificity, phrasing, and the absence of redundancies in the hint. For instance, if (i) a hint covers all mistakes and has harmful extraneous information (-2), and (ii) the hint is not well-phrased and is not specific (-2), then it will be given the score -2-2 = -4. The highest score is 1, denoting a perfect hint that can be used as is, whereas the lowest score is -6 for hints that do not cover any mistakes.
As we progress from left to right at each level, the hint quality decreases by one point, except in cases where harmful extraneous information is present, leading to a two-point reduction. This adjustment was made because harmful extraneous information introduces misleading knowledge, which can be more detrimental to student learning than other issues affecting hint quality. 

Each each student-generated and GPT-generated hint was rated using this rubric. Following this, we found that no hints were rated as -5. Therefore, we finally had 7 different ratings (-6, -4, -3, -2, -1, 0, 1), which we mapped to positive values ranging from 1 to 7, i.e., -6 was mapped to 1, and 1 was mapped to 7.

\subsubsection*{\textbf{Understanding the Dynamics of Hint Writing (RQ2)}}
The first author employed thematic analysis \cite{braun2006using} on all hints generated throughout the study to examine the dynamics of student engagement with the GPT-hints.
After iterative analysis of the data, the first author identified themes in the differences between GPT and student-generated hints.
These themes were then applied to the data collected from each student's engagement in hint writing: (i) whether the GPT-hint introduced any bias, such as extraneous information, redundancies, or revealing the full solution, (ii) whether students identified mistakes that were not identified in the GPT-hint, (iii) whether students disregarded the GPT-hint, for example, by omitting mistakes that were identified in the GPT-hint, and (iv) for the \textit{Deferred AI-assistance} condition, revisions made to their independently written hints. 
Additionally, we recorded the time spent on the hint writing. 
For students in the \textit{AI-assistance} condition, we also captured whether students clicked the ``Show ChatGPT Hint'' button to reveal the GPT-hint.

\subsubsection*{\textbf{Understanding Students' Learning Experience (RQ3)}}
\label{sec:505survey}
In a survey administered to the students after completing both hint writing assignments, we asked them to rate their agreement on a 5-point Likert scale to four statements regarding their experience with hint writing (see Figure \ref{fig:survey}). 
Students in the \textit{AI-assistance} and \textit{Deferred AI-assistance} conditions were also asked to indicate whether the tasks helped them think critically about the responses provided by ChatGPT. 
Additionally, we gave students the option to share any other opinions regarding the hint writing assignments. The first author conducted a thematic analysis of these optional survey responses and categorized them into perceived benefits of the assignments, criticisms, or suggestions for improvement.

\section{Results}
We now present the results for each of the research questions. 
Following convention in psychology studies \cite{pritschet2016marginally}, we consider p<0.05 to be statistically significant, while a p-value between 0.05 and 0.1 to be of marginal significance indicating a trend that is close to statistical significance.

\subsubsection*{\textbf{Sample Statistics}} Out of 101 students enrolled in the course (mean age = 30.94, SD = 8.60), 97 and 95 students participated in the hint writing activities in Week 1 and Week 2, respectively. Table \ref{tab:students_per_cond} shows the number of students, average hint length and time spent (in minutes) on hint writing per condition per week. In contrast with the student-generated hints, the mean (std) lengths of the GPT-hints in Week 1 and Week 2 were respectively: 685.25 (125.35) and 618.62 (168.05). From Table \ref{tab:students_per_cond} we observe that, on average, GPT-generated hints were substantially longer than student-generated hints, reflecting the verbosity and redundancy typical of LLM outputs at the time the study was conducted. The full dataset of student-generated hints, GPT-hints, incorrect solutions for which hints were written along with details of the programming assignments on which the hint writing tasks were based can be found here\footnote{\label{main_footnote}\href{https://osf.io/w538q/?view_only=352e55d06c5543c581b1885702dd4172}{Link to the hints dataset, rubric details, task instructions, and GPT-4 prompts: https://tinyurl.com/25xvsjpn}}.

\subsubsection*{\textbf{Quality of GPT-hints}} For providing AI-assistance, a total of 25 GPT-generated hints of variable quality were generated, corresponding to 12 incorrect solutions for the second programming assignment and 13 for the third (as described in Section \ref{study_design}). These hints were evaluated using the same rubric that was used to evaluate student-generated hints. For Assignment 2, GPT-hints received the following ratings: 1 (n=1), 2 (n=1), 3 (n=3), 4 (n=3), 5 (n=1), and 6 (n=3). For Assignment 3, ratings were similarly distributed as follows: 1 (n=1), 2 (n=1), 3 (n=5), 4 (n=2), 5 (n=0), and 6 (n=4).

\subsubsection*{\textbf{Time on Task}} As expected, students in the \textit{Deferred AI-assistance} condition, who were required to write hints twice within each assignment---both before and after seeing the GPT-hint---spent more time on the entire hint-writing activity than those in the other conditions (see Table \ref{tab:students_per_cond}). This difference was more pronounced in Week 1, when their average time on task was more than double that of the other groups. An ANOVA confirmed that this difference was statistically significant in Week 1 (p = 0.02), but not in Week 2 (p = 0.35). Despite these differences in time on task, we did not include it as a covariate in subsequent models, as our goal was to evaluate the three conditions holistically and capture the outcomes of these designs in authentic implementation contexts.

\begin{table}
\small
\centering
\begin{tabular}{lllll}
\hline
  & Condition & \textit{n} & Hint length & Time spent (mins)\\ \hline 
\multicolumn{1}{l|}{\multirow{4}{*}{1}} & \multicolumn{1}{l|}{Baseline} & 30 & 450.31 (344.41) & 11.13 (6.92--20.21)\\ \cline{2-5}  
\multicolumn{1}{l|}{} & \multicolumn{1}{l|}{AI-assistance} & 39 & 372.18 (244.68) &  11.60 (7.57--22.63)\\ \cline{2-5} 
\multicolumn{1}{l|}{} & \multicolumn{1}{l|}{Deferred AI-assistance} & 28 & 304.70 (202.78)  & 27.68 (18.32--50.67) \\ \hline 
\multicolumn{1}{l|}{\multirow{4}{*}{2}} & \multicolumn{1}{l|}{Baseline} & 28 & 369.50 (329.61)  & 10.12 (4.93--21.13)\\ \cline{2-5} 
\multicolumn{1}{l|}{} & \multicolumn{1}{l|}{AI-assistance} & 38 & 298.18 (206.20) & 10.30 (4.49--18.61)\\ \cline{2-5}
\multicolumn{1}{l|}{} & \multicolumn{1}{l|}{Deferred AI-assistance} & 29 & 245.79 (188.51) & 12.73 (4.93--21.13)\\ \hline
\end{tabular}
\caption{Number of students (\textit{n}), mean (std) of hint length based on number of characters, and median (inter quartile range) of time spent on hint writing per condition in each week--1 \& 2.}
\label{tab:students_per_cond}
\vspace{-5mm}
\end{table}

\subsubsection*{\textbf{Learning Outcomes (RQ1)}}
97 students completed the pre-test, out of which 62 completed the post-test\footnote{Despite the post-test being required, a printing mistake in the syllabus caused some to infer that it was optional, leading to missing post-test data for 35 students.}. Out of these 62 students, 20, 27, and 15 students were respectively from the \textit{Baseline}, \textit{AI-assistance}, and \textit{Deferred AI-assistance} conditions. An ANOVA revealed that the difference in pre-test scores of these students between conditions was marginally significant (p = 0.09), with the \textit{AI-assistance} group receiving the lowest pre-test scores on average.  Although this difference in pre-test scores did not reach statistical significance, given the small sample size, we adopted a conservative approach and used propensity score matching \cite{abadie2016matching} to select a sample of 20 students from the \textit{AI-assistance} group whose pre-test scores were comparable to the average pre-test scores of the other two groups. Following this, we had 20, 20, and 15 students in the \textit{Baseline}, \textit{AI-assistance}, and \textit{Deferred AI-assistance} conditions. Table \ref{tab:scores} shows the aggregates of pre- and post-test scores for this group of 55 students. An ANOVA confirmed that there was no significant difference in their pre-test scores (p = 0.82). On comparing the post-test scores of these students who had similar levels of prior knowledge, we found that students in the \textit{AI-assistance} condition had the lowest average post-test score of 4.75, while students in the \textit{Baseline} and \textit{Deferred AI-assistance} conditions had higher average post-test scores of 5.70 and 5.67 respectively. However, an ANOVA revealed that the difference in post-test scores of students between conditions was not statistically significant (p = 0.18). 

\begin{table}
    \centering
    \begin{tabular}{c|c|c|c}
    \hline
          Condition&  \# students&  Pre-test & Post-test\\
          & & (mean, std) & (mean, std)\\
\hline
 Baseline&  20&  (7.80, 1.40)& (5.70, 1.81)\\
AI-assistance&  20&  (7.55, 1.05)& (4.75, 1.65)\\
Deferred AI-assistance&  15&  (7.60, 1.55)& (5.67, 1.84)\\
\hline
    \end{tabular}
    \caption{Number of students and aggregates of pre- and post-test scores for each experimental condition. For each test, the maximum score was 10.}
    \label{tab:scores}
    \vspace{-5mm}
\end{table}

\subsubsection*{\textbf{Impact of Hint Writing Design on Hint Quality: Quantitative Findings (RQ2)}}
We used a linear mixed-effects model with the quality rating of the final student-generated hints as the dependent variable. The experimental condition, week number, and the number of mistakes in the incorrect solution were used as fixed effects, and student ID was used as a random effect. The variable representing the number of mistakes in the incorrect solution was assigned a value of 1 if there was a single mistake and 2 if there were multiple mistakes. This approach aligns with the rubric (Figure \ref{fig:utility_levels}), which differentiates between hints that address all mistakes in the incorrect solution and those that only address some.

We found that, compared to the \textit{Baseline} condition, students in the \textit{Deferred AI-assistance} condition ($\hat{\beta}$=1.096, SE=0.37, p=0.003) and the \textit{AI-assistance} condition ($\hat{\beta}$ = 0.75, SE=0.34, p=0.03) wrote higher quality hints. Post-hoc pairwise comparisons using the Tukey method (for family-wise error rate adjustment) revealed that the quality of student-generated hints in the \textit{Deferred AI-assistance} condition was significantly better than those in the \textit{Baseline} condition (p = 0.01), and the difference between the \textit{AI-assistance} and \textit{Baseline} conditions was marginally significant (p = 0.08). 
We also compared hints from the three conditions based on coverage of mistakes and the exclusion of extraneous information using the Level 1 score from the rubric (see Figure \ref{fig:utility_levels}). We used the same linear mixed effects model but used the Level 1 score for the hint as the dependent variable. Post-hoc pairwise comparisons using the Tukey method revealed that students in the \textit{Deferred AI-assistance} condition were significantly more successful than those in the \textit{Baseline} condition at identifying mistakes and excluding extraneous information (p=0.03). There was no significant difference between the other two pairs of conditions. 

\begin{figure}[t]
  \centering  \includegraphics[width=\columnwidth]{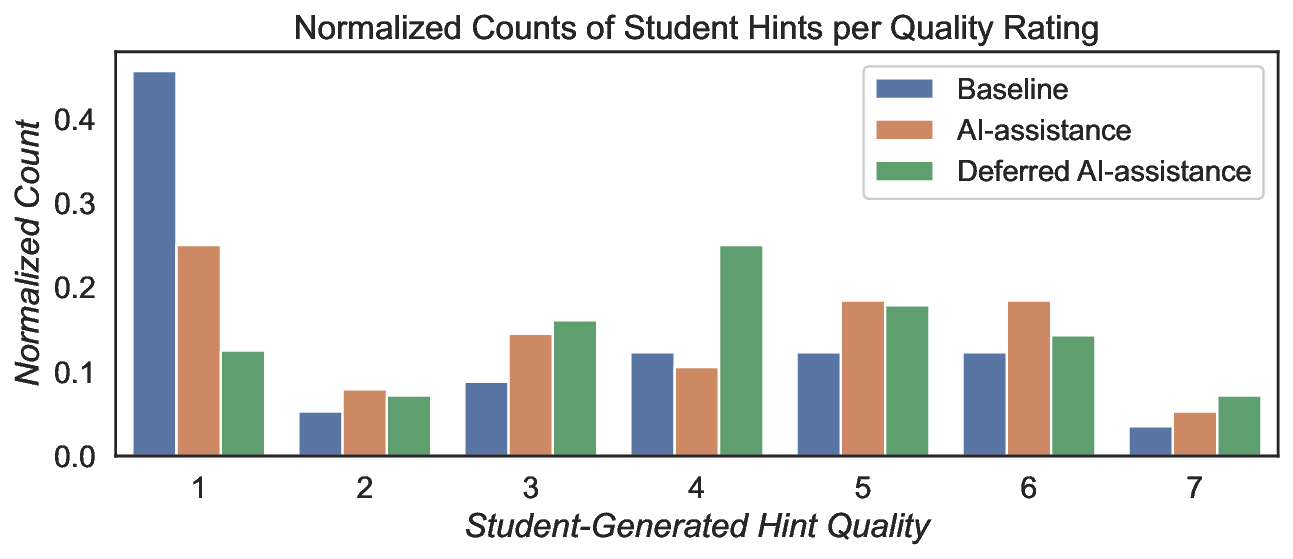}
  \caption{Barplots depicting normalized counts of student-generated hints per quality rating (7 being the highest) for each condition.} 
  \vspace{-4mm}
    \label{fig:505hint_qual_distribution}
\end{figure}

Overall, 
the \textit{Deferred AI-assistance} design led to greater success in writing high-quality hints, identifying mistakes in the incorrect solution, and excluding extraneous information. Figure \ref{fig:505hint_qual_distribution} shows, for each condition, the distribution of normalized counts of student-generated hints per hint-quality rating.

\subsubsection*{\textbf{Impact of Hint Writing Design on Hint Quality: Qualitative Findings (RQ2)}}
We now provide the results of the qualitative analyses regarding the hints written by students in each condition.
Around 45\% of the hints generated by students in the \textit{Baseline} condition did not identify any of the mistakes in the incorrect solutions, compared to 12.5\% and 25\% from the \textit{Deferred AI-assistance} and \textit{AI-assistance} conditions.
Amongst such hints from the \textit{Baseline} condition, some were vague, provided surface-level guidance, and did not identify some trivial mistakes in the incorrect solution. 
These patterns suggest that these students in the \textit{Baseline} condition did not meaningfully engage in hint writing, but rather wrote the hints perfunctorily to receive the points awarded for task completion. 

Turning to the hints written by students in the \textit{AI-assistance} condition, 
for around 34\% of the hints, we did not find any negative influence of the GPT-hint. 
On the other hand, for around 17\% of the hints, we found evidence for negative impacts of GPT-hints, such as the presence of extraneous information, redundancies, or giving away the full solution. Few students simply reworded the GPT-hint.
Approximately 20\% of students in the \textit{AI-assistance} condition did not view the GPT-hint, and 17\% appeared to disregard the GPT-hint despite viewing it, as evidenced by their failure to identify mistakes that had been correctly explained in the GPT-hint. Among these students, some produced low-quality hints, many of which failed to identify any mistakes, suggesting a lack of engagement with the task. However, there were also those who disregarded the GPT-hints yet still produced high-quality hints.

Next, we analyzed the hints written by students in the \textit{Deferred AI-assistance} condition.
Except for one student in Week 1 and four students in Week 2, all students in this condition made revisions to the independently-written hints after viewing the GPT-hint.
Around 52\% students in the \textit{Deferred AI-assistance} condition made improvements to their original hint. Many of these students were able to identify more mistakes in the revised hint. Some students were able to remove harmful extraneous information from their original hints to write better revised hints. Several students improved the phrasing of their hints, possibly because of the influence of the GPT-hints that were typically well phrased. Some students were also able to make their hints more specific. 
For instance, a student originally wrote a hint in which one of the mistakes was not immediately clear and later revised it by improving the phrasing and making it specific: 
\begin{quote}
\textbf{Student's original hint}: ``\textit{The denominator must contain the number of cases where the person was vaccinated but did not contract chicken pox. This code also includes cases where chicken pox was caught.}". 
\end{quote}
\begin{center}
    \begin{tikzpicture}
        \draw[->, thick] (0,0) -- (0,-0.35);
    \end{tikzpicture}
\end{center}
\begin{quote}
\textbf{Student's revised hint}: ``\textit{Focus on correctly filtering vaccinated children who contracted chickenpox versus those who were vaccinated but did not contract chickenpox in your calculations. Ensure that your denominator includes only cases where individuals were vaccinated but did not contract chickenpox...}" 
\end{quote}

\noindent For around 32\% of students in the \textit{Deferred AI-assistance} condition, there was no major difference between the quality of their revised hints and the original hints. Around 39\% of students made revisions that negatively affected the quality of the revised hint, occasionally alongside improvements.
For instance, some students wrote revised hints that gave away the full solution, even when they originally wrote hints that did not do so, as evidenced by the following:
\begin{quote}
\textbf{Original hint}: ``\textit{Are you sure you are taking the right column? What does the \texttt{VRC1\_AGE} column represent?}"
\end{quote}
\begin{center}
    \begin{tikzpicture}
        \draw[->, thick] (0,0) -- (0,-0.35);
    \end{tikzpicture}
\end{center}
\begin{quote}
\textbf{Revised hint}: ``\textit{The \texttt{VRC1\_AGE} column returns the age at which they received. You should consider using the \texttt{P\_NUMVRC} column, which represents the total number of varicella doses the child receives...}"
\end{quote}

\noindent Some students added extraneous information or redundancies to their revised hints after seeing the GPT-hint. One student removed a fix for a mistake that they had originally included in their hint, possibly because the GPT-hint did not identify that mistake. 


Finally, we noted that students in both the \textit{AI-assistance} and \textit{Deferred AI-assistance} conditions struggled to identify mistakes that were not identified in the GPT-hint. Most of these mistakes were logical mistakes that were challenging to identify. However, some mistakes were trivial, such as rounding the output to 5 decimal places despite being instructed not to do so. Yet, most of the students in both the AI assistance conditions were unable to identify this mistake, suggesting potential over-reliance on the GPT-hint.

Overall, the quantitative and qualitative analyses regarding hint quality revealed two main insights. First, students in the \textit{Baseline} condition often struggled to identify mistakes or engaged with the task superficially. Second, while students in the \textit{Deferred AI-assistance} condition were most successful in achieving broad coverage of mistakes, they were also relatively more susceptible to negative influences from GPT-generated hints in terms of effectiveness of communication---for instance, by disclosing full solutions or introducing redundant information in their final hints.

\subsubsection*{\textbf{Student Perceptions of the Hint Writing Tasks (RQ3)}}
\begin{figure}
  \centering  \includegraphics[width=\columnwidth]{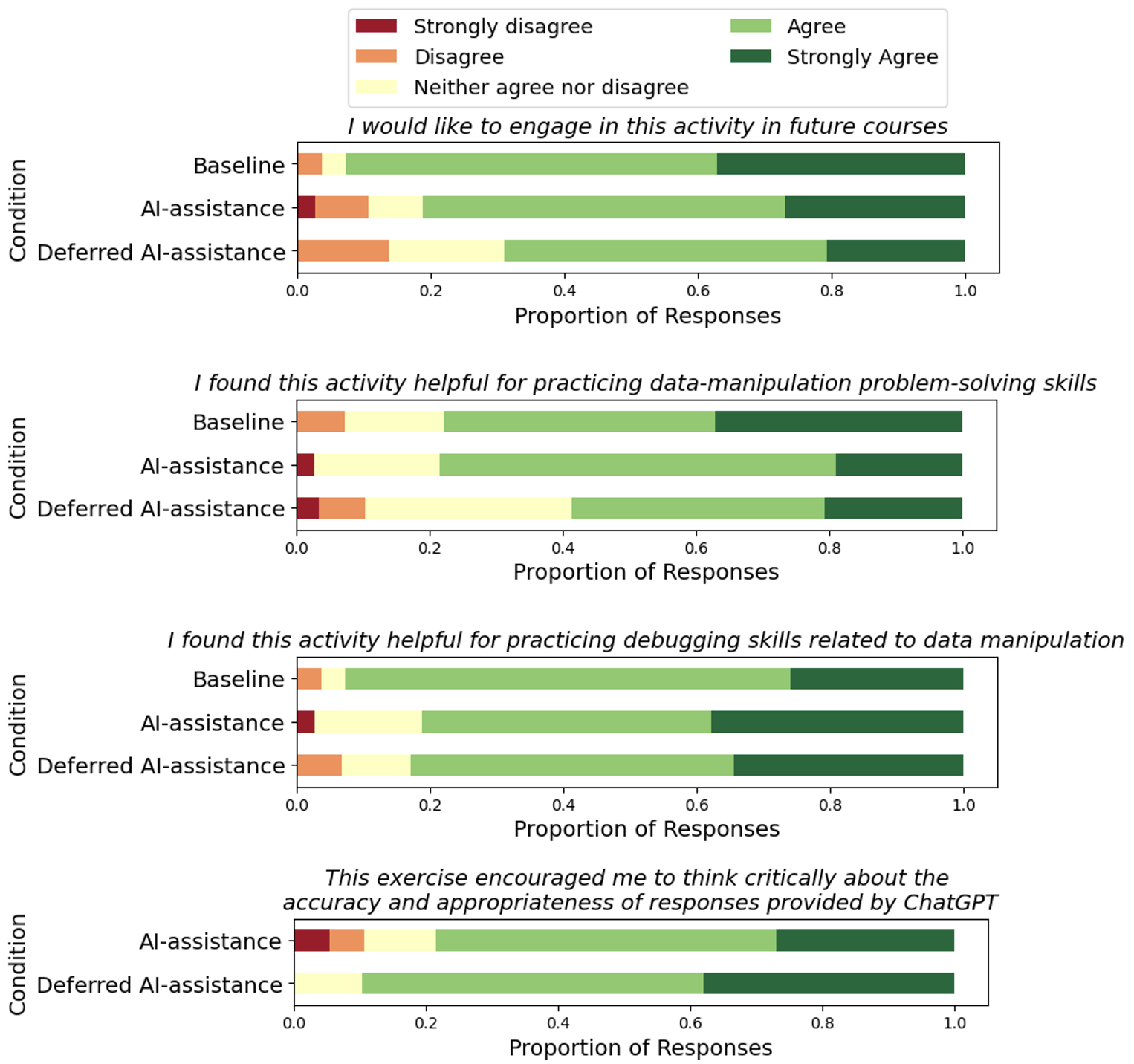}
  \caption{Distribution of responses for the survey questions examining learners' perceptions of the hint writing tasks.} 
    \label{fig:survey}
    \vspace{-3mm}
\end{figure}

Students' perceptions of the hint writing tasks were generally positive, indicating a broad appreciation and acceptance of these tasks (Figure \ref{fig:survey}). 
A majority of students expressed willingness to participate in hint writing activities in future courses, with those in the \textit{Baseline} condition showing the most enthusiasm, followed by those in the \textit{AI-assistance} condition. Further, a majority of the students across all conditions found the activities helpful for practicing debugging skills. 
A smaller proportion of students in the \textit{Deferred AI-assistance} condition found the activity useful for practicing data manipulation skills compared to the other two conditions, likely because the task design directed their attention more toward evaluating the AI hint than toward domain-specific data manipulation skills.
For the fourth survey statement on helpfulness of the exercise for thinking critically about the GPT-generated responses, the majority of the students in both the AI support conditions were in agreement. However, around 10\% of the students from the \textit{AI-assistance} condition were not in agreement. 

Analysis of students' open-ended survey responses regarding their perceptions of the hint writing tasks revealed that many students appreciated the opportunity to review a (hypothetical) peer's work as it provided exposure to alternative ways of thinking and solving a problem (``\textit{I like these [hint-writing] activities, really make me understand how different ways can the same code be written.}'', ``\textit{I've never [engaged in an activity] like this, but I appreciated the opportunity to do a code review and practice critical thinking.}''
). One student noted their appreciation for the assignment's focus on writing and communication, highlighting it as a welcome contrast to the predominantly programming-focused activities in the course. Overall, all (except one) students who responded to the open-ended survey question found the task helpful for learning (``\textit{...as uncomfortable/challenging as it is, I believe it helps me `think' like a data scientist more, which is positive.}'', ``\textit{It puts you in the shoes of a teacher, so it will definitely enhance the debugging skills.}''). Some students expressed their interest in providing feedback directly to their peers through such hint writing activities, suggesting their openness to integrating such approaches into peer feedback settings where they can exchange hints with one another (``\textit{...I thought this was a good activity for giving feedback to people in the future.}'').

Turning to the critiques of the hint writing exercises and opportunities for improvement, one student was unsure about the utility of the hints they wrote for helping a peer fix their code (``\textit{I think this is a good assignment, but it would be useful to actually write out the solution and detail the steps. The hints can be too vague if the solution is wildly incorrect or inefficient. Each little [fix] leads to another set of changes.}''). A few students felt that this assignment was designed to collect training data for fine-tuning LLMs, which was not the case. 


\section{Discussion and Future Work}
We now discuss the study’s findings and outline directions for future research, focusing on the impact of each hint-writing design on student learning and hint quality, strategies for mitigating the negative influence of generative AI assistance and students' perceptions of the pedagogical value of the hint writing activities.

\subsubsection*{\textbf{Learning Outcomes \& Engagement}}
RQ1 probed the impact of each hint writing design on students' learning outcomes. Although we observed a trend indicating lower learning outcomes for students in the \textit{AI-assistance} condition compared to other conditions, this difference was not statistically significant. 
A potential reason for this could be the limited sample size for measuring differences in learning outcomes (n=55) and the relatively small weight of the hint-writing intervention compared to other course activities. Future research with a larger sample is needed to investigate whether requiring students to independently attempt solutions prior to engaging with AI support leads to improved learning outcomes compared to providing on-demand AI assistance.

 

The overall positive student response to the hint writing tasks (RQ3) confirms that it is a valuable addition to their curriculum, encouraging them to think deeply about the course material and AI-generated responses. Further, the automated pipeline for generating  personalized hint writing tasks (described in Section \ref{study_design}), where students can compare their correct solution with incorrect solutions from peers, offers a scalable method for incorporating reflective hint writing into large courses. 
As we did not provide feedback to the students regarding the quality of their hints, future research should also explore how to provide such feedback at scale, using the provided rubric and dataset resulting from this work.

\subsubsection*{\textbf{Deferred AI-assistance to Foster Independent Thinking}}
\label{dis:quality}
Probing into the impact of the hint writing designs on the quality of student-generated hints (RQ2(a)), we found that AI support helped students write higher quality hints, and this result was statistically significant for the \textit{Deferred AI-assistance} condition. In the absence of any AI support, a high proportion of students from the \textit{Baseline} condition wrote hints which did not accurately identify any mistakes. Further, deferring AI assistance led to significantly higher success rate of writing of hints that were accurate and excluded extraneous information. 

Notably, students in the \textit{Deferred AI-assistance} condition spent substantially more time on the hint-writing activity in Week 1---on average, more than twice as long as students in the other conditions. In Week 2, however, the between-condition differences in time on task were not significant. Two observations emerge from this pattern. First, it is expected that students asked to write two hints, as in the \textit{Deferred AI-assistance} condition, may spend roughly twice as much time as those writing only one hint. Yet, the fact that their time exceeded this threshold and that they also had the highest success in identifying mistakes suggests that this design enhanced cognitive engagement in a desirable way. As demonstrated in prior research \cite{shibani2024untangling}, learners often exhibit shallow engagement with LLMs unless explicitly prompted to engage in self-reflection, which could explain the lower time on task of the \textit{AI-assistance} condition. 
Deferring AI assistance 
could have led to students engaging in active reflection and revision of their original hint, as well as more critical evaluation of the AI-generated hint. This is corroborated by recent research \cite{steinbach2025llms} which shows that LLM-generated feedback, even when consisting of hallucinations, can enhance learning by prompting error detection, which in turn fosters greater cognitive engagement.
Second, the disappearance of any significant difference in time on task in Week 2 could be due to several factors such as novelty effect in Week 1 or that the students became more familiar with the task in Week 2 and, as a result, invested less cognitive effort. They may also have adopted more efficient strategies in Week 2, such as spending less time writing they first hint before viewing the GPT hint, although we could not measure this time separately. More research is needed to disentangle these hypotheses and to explore alternative forms of student-AI collaboration that sustain student engagement over time. 

\subsubsection*{\textbf{Enhancing AI-Assistance for Learning from Hint Writing}}
Around 17\% students in the \textit{AI-assistance} condition and 39\% students in the \textit{Deferred AI-assistance} condition wrote hints that showed signs of being negatively impacted by the GPT-hints provided for assistance, e.g., including redundant or extraneous information. Importantly, these effects in the \textit{Deferred AI-assistance} condition were concentrated in the communicative quality of students’ hint rather than in coverage of mistakes, for which the this design was actually most effective. The mandatory exposure and directive to compare one’s independently written hint with the GPT-hint in the \textit{Deferred AI-assistance} condition likely increased the cognitive load of the task and amplified the negative influences of the GPT-hints: these students were instructed to revise their hints based on the GPT-hint, whereas students in the AI-assistance condition could consult the GPT-hint at their discretion. Since this experiment was conducted in 2023 and LLM capabilities have evolved substantially since then, such communicative harms are not necessarily inevitable and can be mitigated through better model selection and LLM prompting techniques. For instance, prompt engineering can be improved to ensure that the AI-generated hints have no redundancies and contain minimal extraneous information, such as using chain-of-thought prompting or prompt chaining \cite{wei2022chain}. 
However, if the primary goal is to provide student hint-writers with a meaningful learning experience, it is worth exploring alternative forms of deferred AI assistance that balance reducing over-reliance with maintaining appropriate cognitive load. For example, Vasconcelos et al. \cite{vasconcelos2023generation} found that confidence-based highlighting of potential inaccuracies in AI-generated code—based on estimated token modifications or deletions—improved programmer productivity. This approach can be useful for providing an alternative form of assistance in hint-writing, where students first generate hints independently and then receive highlighted inaccuracies in the incorrect solution rather than a potentially inaccurate AI-generated hint. Compared to the \textit{Deferred AI-assistance} design we explored, which required both error identification and reflection on an AI-generated hint, this approach can help sustain engagement by reducing unnecessary cognitive demands while preserving critical learning benefits and focusing on the core learning objective of error identification.

\subsubsection*{\textbf{Misalignment Between Student Perceptions and Hint Writing Outcomes}}
Survey responses regarding the hint-writing tasks (RQ3) revealed some discrepancies between students’ perceptions of the task designs and the observed effects. Students in the \textit{Baseline} condition reported the highest willingness to engage in this activity in future courses and also valued it the most for practicing debugging. Yet, this same group also produced the largest proportion of hints that failed to identify a single error correctly (see Figure \ref{fig:505hint_qual_distribution}). These students engaged entirely independently in hint writing, and while this may have fostered some metacognitive engagement \cite{fisher2021harder}, it did not translate into success in accurately diagnosing errors.

In contrast, the \textit{Deferred AI-assistance} condition was perceived relatively less positively, a finding that may be attributed to students’ unfamiliarity with the approach and the additional complexity introduced by this design. This is not surprising, as learners often favor minimizing effort \cite{chen2025more}.
Additional factors may have shaped these perceptions such as limited trust in AI-based support \cite{singh2023bridging} and skepticism among few students who viewed the activity as contributing training data. The mixed reactions to LLM use in education documented in prior research \cite{chan2023students, padiyath2024insights} would also have likely influenced student evaluations of the designs with AI assistance. Notably, because the study was conducted in 2023, when AI integration in education was lower and AI literacy was perceived as less important than it is today, students' perceptions---and potentially their intrinsic motivation and outcomes such as hint quality---would likely be more positive if the study were conducted now.

Taken together, these findings highlight both the importance and the challenges of designing student-AI collaborative learning activities with an optimal cognitive load that provide opportunities for productive cognitive engagement. Moreover, the observed over-reliance on AI-generated hints, as evident from students' failure to identify errors that were not captured by AI, underscores the pressing need for pedagogical designs that engage learners in critical, reflective interactions with AI.

\subsubsection*{\textbf{Limitations}}
This study has some limitations, particularly due to being conducted in an authentic educational setting. 
The outcomes of students' engagement in the hint writing tasks could have been impacted by the quality of the AI-generated hints they received. While we did not examine this effect due to the limited sample size, future research with a larger sample of students and AI-generated scaffolds of varying quality can be helpful to address this. Next, we were unable to obtain post-test scores for all students, reducing the study's statistical power for measuring learning outcomes. Future studies with larger sample sizes involving multiple iterations of student-AI collaborative learning activities
could be useful for understanding the 
impacts on student learning. Next, an important finding of our study was that students in the \textit{Deferred-AI assistance} condition spent significantly more time on the hint writing task in the first week compared to those in the other conditions. While this additional time likely contributed to improved hint quality, our primary focus was to understand how different hint writing designs influence hint quality, learning outcomes, and engagement, independent of time on task. Given this limitation, future research should explore the extent to which time on task, independent of task design, influences hint quality and learning outcomes.
Finally, we focused on a single course at a single institution, although including two hint writing interventions based on different programming assignments mitigated this limitation to some extent. 

\section{Conclusion}
This work contributes an algorithmic approach for developing personalized and reflective hint writing assignments for data science learners, and investigates the effects of three hint writing designs---two of which involve student-AI collaboration---on the quality of the student generated hints, learning outcomes and student perceptions. Through mixed-methods analysis of students' engagement in these activities, we show that offering AI assistance \textit{after} students have independently attempted to write a hint results in higher-quality hints and leads to greater success in error identification. Given the growing concerns around cognitive offloading in student-AI interactions, deferring AI assistance presents a simple and scalable approach to reducing over-reliance on LLMs while also fostering students' independent thinking. 
While students generally viewed the hint-writing activities positively, the somewhat lower perceived value of deferred AI assistance, compared to on-demand or no assistance, highlights the tension between fostering optimal cognitive engagement and maintaining student motivation. To address this challenge, we outlined future research directions aimed at developing alternative forms of student-AI collaboration that sustain engagement while mitigating the risk of over-reliance. Overall, this work contributes to research on enhancing student learning in the age of AI through meaningful student-AI interactions.



\bibliographystyle{ACM-Reference-Format}
\bibliography{bibliography}

\appendix

\end{document}